\documentclass[prl,twocolumn,citeautoscript,superscriptaddress]{revtex4}

\usepackage[dvips]{graphicx}
\usepackage{epsfig}
\usepackage{dcolumn}
\usepackage{amsmath}

\newcommand*{\Et}{\mathcal{E}}

\usepackage{times}

\begin{document}
\title{Hyperfine tensors of nitrogen-vacancy center in
diamond from \emph{ab initio} calculations}

\author{Adam Gali} 
\affiliation{Department of Atomic Physics, Budapest
  University of Technology and Economics, Budafoki \'ut 8., H-1111, 
  Budapest, Hungary}
\affiliation{Department of Physics and School of 
Engineering and Applied Sciences, Harvard University, Cambridge, Massachusetts 02138, USA}

%\author{Efthimios Kaxiras} \affiliation{Department of Physics and
%  School of Engineering and Applied Sciences, Harvard University,
%  Cambridge, Massachusetts 02138, USA}

\begin{abstract}
We determine and analyze the charge and spin density distributions of
nitrogen-vacancy (N-V) center in diamond for both the ground and
excited states by \emph{ab initio} supercell calculations. We show
that the hyperfine tensor of $^{15}$N nuclear spin is negative and
strongly anisotropic in the excited state, in contrast to previous
models used extensively to explain electron spin resonance
measurements. In addition, we detect a significant redistribution of
the spin density due to excitation that has serious implications for
the quantum register applications of N-V center.
\end{abstract}

\maketitle

%\section{Introduction}

Nitrogen-vacancy (N-V) centers in diamond have numerous peculiar
properties that make them a very attractive solid state system for
fundamental investigations of spin based phenomena. Recently, this
defect has been proposed for several applications, like quantum
information processing \cite{Gruber97,Epstein05,Childress06,Hanson08},
ultrasensitive magnetometer \cite{Maze08,Balasubramanian08}, and
measurement of zero-point fluctuations or preparing quantum-correlated
spin-states over macroscopic distance \cite{wrachtrup09}. In these
measurements, a room temperature read-out of \emph{single nuclear
spins} in diamond has been achieved by coherently mapping nuclear spin
states onto the electron spin of a single N-V center
\cite{Jelezko04-2,Childress06} which can be \emph{optically} polarized
and read out with long coherence time
\cite{Jelezko04-1,Gaebel06}. Particularly, this has been the basis in
realization of a nuclear-spin-based quantum register
\cite{Childress07} and multipartite entanglement among single spins at
room temperature \cite{Neumann08}. The polarization of a single
nuclear spin has been achieved by using either a combination of
selective microwave excitation and controlled Larmor precession of the
nuclear-spin state \cite{Childress07} or a level anticrossing \emph{in
the excited state} \cite{jacques:057403}. Understanding the spin
states and levels is of critical importance for optical control of N-V
centers in both the ground and excited states. Especially, the
hyperfine interaction couples the electron spin and nuclear spin, thus
determination of hyperfine tensors of the nuclei with non-zero nuclear
spin plays a key role both in creation of entaglement states and in
the decoherence process \cite{Childress06,maze:094303}.

%\section{Results briefly}

Recently, the hyperfine signals in the ground
\cite{jacques:057403,felton:075203} and excited
\cite{fuchs:117601} states have been detected in N-V centers but with
contradicting interpretations. In a conventional electron paramagnetic
resonance (EPR) spectrum on the ensemble of N-V centers, the $^{15}$N
signal was assumed to be \emph{positive} with slight anisotropy in the
ground state while Fuchs \emph{et al.} and Jacques \emph{et al.}
assumed an isotropic \emph{negative} hyperfine constant for $^{15}$N
in the ground state \cite{fuchs:117601,jacques:057403} based on
previous EPR and optically detected magnetic resonance (ODMR)
measurements \cite{He93,rabeau:023113}. Recently, Fuchs \emph{et al.}
have reported that the hyperfine splitting of $^{15}$N should be
$\sim$20$\times$ larger in the excited state than in the ground state
\cite{fuchs:117601}. The excited-state $^{15}$N hyperfine signal was
assumed to be isotropic in their model \cite{fuchs:117601}. While we
already addressed the hyperfine tensor of $^{14}$N in the ground state
\cite{Gali08}, the lack of the detailed study on $^{15}$N hyperfine
signal and the proximate $^{13}$C isotopes \emph{both} in the ground
and excited states prohibits the understanding of the intriguing
physical properties of this defect. In this Letter, we thoroughly
investigate the hyperfine tensors of proximate $^{15}$N and $^{13}$C
isotopes of the N-V center \emph{both} in the ground and excited
states by means of high level \emph{ab initio} supercell all-electron
plane wave calculations. In addition, we analyse the overall charge
and spin density distributions before and after the optical
excitation. 
%We have found the following effects of optical excitation:
%i) considerable re-arrangment of atoms near the vacant site ii) charge
%density redistribution iii) large spin density redistribution. 
We show that the hyperfine constants of $^{15}$N is \emph{positive}
and possibly slightly anisotropic in the ground state while
\emph{negative} and \emph{strongly anisotropic} in the excited
state. In addition, the hyperfine splittings of the proximate $^{13}$C
isotopes change also significantly that has serious implications both
in the interpretation of the excited-state spectroscopy signals and in
the quantum-information applications.

%\section{Model and Computational methodology}

\begin{figure}
%\begin{center}
%\includegraphics[keepaspectratio,totalheight=6cm,width=8.0cm]{Figure1.eps}
\includegraphics[width=8.0cm]{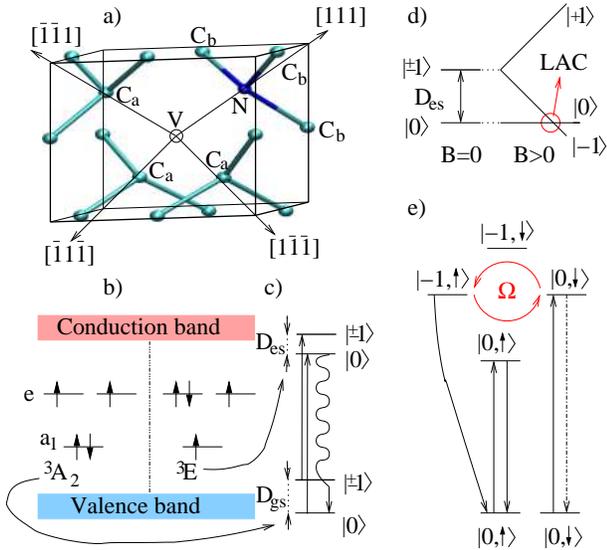}
%\end{center}
\caption{\label{fig:structlev}(Color online) a) The structure of
  nitrogen-vacancy (N-V) defect in our particular working frame. The
  place of vacancy is denoted by an empty circle. b) The schematic
  single particle picture of the $m_s=1$ high spin states in ground
  state ($^3A_2$,gs) and excited state ($^3E$,es). c) The fine
  structure of the $^3A_2$ and $^3E$ states at room temperature due to
  spin-spin interaction. Zero-field splittings are
  D$_\text{gs}$=2.88~GHz \cite{Loubser77}, D$_\text{gs}$=1.42~GHz
  \cite{Batalov09}. During optical excitations the fluorescence is
  predominantly active for the $m_s=0$ ground state ($|0\rangle$)
  due to a non-radiative intersystem crossing of the $|\pm1\rangle$
  es states with the many-body singlet states (not shown here). d)
  Splittings of the es substates in the presence of the magnetic field
  (B). LAC is expected between $|0\rangle$ and $|-1\rangle$ states. e)
  Simplified energy-level diagramm with including the hyperfine
  structure associated with $^{15}$N nuclear spin states
  $|\uparrow\rangle$ and $|\downarrow\rangle$ in the case of LAC
  regime of the applied B-field. At LAC, precession frequency $\Omega$
  between excited-state sublevels $|0,\downarrow\rangle$ and
  $|-1,\uparrow\rangle$ can lead to nuclear-spin flip, which can be
  transfered to the ground state through nonradiative intersystem
  crossing (curved arrow).}
\end{figure}

The negatively charged NV-center in diamond \cite{Davies76} consists
of a substitutional nitrogen atom associated with a vacancy at an
adjacent lattice site (Fig.~\ref{fig:structlev}a). The ground state
has $^3A_2$ symmetry where one $a_1$ defect level in the gap is fully
occupied by electrons while the double degenerate $e$-level above that
is occupied by only two electrons with parallel aligment of spins
(Fig.~\ref{fig:structlev}b). Thus, this defect has S=1 high spin
ground state \cite{Goss96}. By promotion one electron from the
$a_1$-level to the $e$-level will result in the excited $^3E$
state. Both states can be described by conventional density functional
methods. The ground state can be readily described by spin density
functional theory while the excited state can be obtained by
constrained occupation of states \cite{Gali08}. We optimized the
geometry for both the ground and excited states, and calculated the
charge and spin densities at the optimized geometries.

We applied the PBE functional \cite{PBE} to calculate the spin density
of the defect. First, the diamond primitive lattice was optimized,
then a simple cubic 512-atom supercell was constructed from
that. Finally, we placed the negatively charged NV-defect into the
supercell, and optimized under the given electronic configuration. We
utilized the {\sc VASP} code for geometry optimization
\cite{Kresse96}. We applied plane wave basis set (cut-off: 420~eV)
with PAW-method \cite{Blochl94}.  During the optimization of the
lattice constant we applied twice as high plane wave cut-off and
12$^3$ Monkhorst--Pack K-point set \cite{MP76}. For the 512-atom
supercell we used the $\Gamma$-point that provided convergent charge
density. We plugged the optimized geometry into the {\sc CPPAW}
supercell plane wave code with PAW-method that provides the hyperfine
tensors \cite{CPPAW}. We applied the same basis set and projectors in
both codes yielding virtually equivalent spin density of the
defects. Other technical details are given in
Ref.~\onlinecite{Gali08}. The charge density distribution was analyzed
by the Bader-method \cite{baderanalyzis}. We briefly mention here that
we provide the principal values of the hyperfine tensors, called,
$A_{11}$, $A_{22}$, and $A_{33}$ that can be found by diagonalization
of the hyperfine tensors. If the hyperfine field has C$_{3v}$ symmetry
then $A_{11}$=$A_{22}$=$A_{\perp}$ and $A_{33}$=$A_{||}$ where $||$
means that hyperfine field coincides with the symmetry axis of the
defect. The Fermi-contact term ($a$) is defined as
$a=(A_{||}+2A_{\perp})/3$ while the dipole-dipole term ($b$) as
$b=(A_{||}-A_{\perp})/3$. The hyperfine field is isotropic when $b=0$,
i.e., $A_{\perp}$=$A_{||}$.

%\section{The calculated geometry, charge density, hyperfine tensors}

First, we discuss the geometry of the defect. The obtained distance
between the carbon atoms is 1.54~\AA\ in perfect diamond. In the
nitrogen-vacancy defect there are three carbon atoms (C$_a$) and one
nitrogen atom (N) closest to the vacant site each possessing a
dangling bond (Fig.~\ref{fig:structlev}a). 
%The defect has C$_{3v}$
%symmetry. 
%Since the second neigbor atoms are relatively far from
%each other it is more favorable for these atoms to relax outward from
%the vacant site than to relax inward to form long bonds. 
The defect conserves its C$_{3v}$ symmetry during the outward
relaxation, and the dangling bonds of the C$_a$ and N atoms will point
to the vacant site. The symmetry axis will go through the vacant site
and the N-atom which is the [111] direction in our particular working
frame (see Fig.~\ref{fig:structlev}a).
%It is important to
%notice that there are four symmetrically equivalent
%$\langle$111$\rangle$ directions, therefore, the individual NV centers
%can have four different orientations with respect to the [111]
%direction. 
We found in our PBE calculations that the C$_a$ atoms are closer to
the vacant site (1.64~\AA ) than the N-atom (1.69~\AA ) in the ground
state. We obtained 5.97$e$ Bader-charge on the N-atom that it is 0.97$e$
larger than the number of valence electrons of the neutral N-atom. 
%At
%the first sight, one may assume that the extra negative charge in the
%defect is accumulated by N-atom. 
N-atom is more electro-negative than
the C-atom. Indeed, the three C-atoms bound to N-atom
(C$_b$) have 3.69$e$ Bader-charge, so there is significant charge
transfer from C$_b$ atoms toward the N-atom ($\sim$0.93$e$ as
total). That is the main source of the negative polarization of the
N-atom. We found that the negative charge is distributed on many atoms
around the defect. The C$_a$ atoms are even slightly positively
polarized (3.97$e$) that will finally induce a dipole moment in the
defect. Next, we briefly discuss the spin density of the ground
state. The spin density is primarily originated from the unpaired
electrons on the $e$ defect level in the gap. Due to symmetry reasons
\cite{Lenef96,Gali08} the $e$-level is only localized on the C$_a$
atoms but \emph{not} on the N-atom. Therefore, large spin density is
expected on the C$_a$ atoms while negligible on the N-atom. Indeed,
very small hyperfine splitting was found for $^{14}$N \cite{He93} and
$^{15}$N \cite{rabeau:023113,felton:075203}. However, the sign of the
Fermi-contact term of the hyperfine interaction for $^{15}$N was
contradictory. Rabeau \emph{et al.} assumed a negative value
\cite{rabeau:023113} while Felton \emph{et al.} has recently proposed
a positive value \cite{felton:075203}. The gyromagnetic factor of
$^{15}$N ($\gamma _\text{N}$) is negative, thus negative(positive)
Fermi-contact hyperfine value ($a$) indicates positive(negative) spin
density on the N-atom ($n_s$) because $a \sim
\gamma _\text{N} n_s$. In our previous LDA calculation \cite{Gali08}
we detected negative spin density on N-atom. Our improved PBE
calculation justifies this scenario. Due to symmetry reasons the
direct spin polarization of N-atom does not occur in the ground state
but the large spin density on the C$_a$ atoms can polarize the core
electrons of the N-atom, i.e., it is an indirect and weak spin
polarization. In our PBE calculation we can also detect a slight
dipole-dipole interaction for $^{15}$N which is outside of our error
bar ($\sim$0.3~MHz \cite{Gali08}). Our conclusions agree with the
findings of Felton \emph{et al.}  \cite{felton:075203}: the $^{15}$N
has \emph{positive} hyperfine splittings and it is slightly
anisotropic (see Table~\ref{tab:hyperf}). The calculated hyperfine
tensors for C$_a$ atoms agree nicely with the recent experimental
values recorded at low temperature \cite{felton:075203}. The fair
qualitative agreement between the experiment and theory for the ground
state allows us to study the less known excited state with our tools.

\begin{figure}
%\begin{center}
%\includegraphics[keepaspectratio,totalheight=6cm,width=8.0cm]{NVgeometel.eps}
\includegraphics[width=7.5cm]{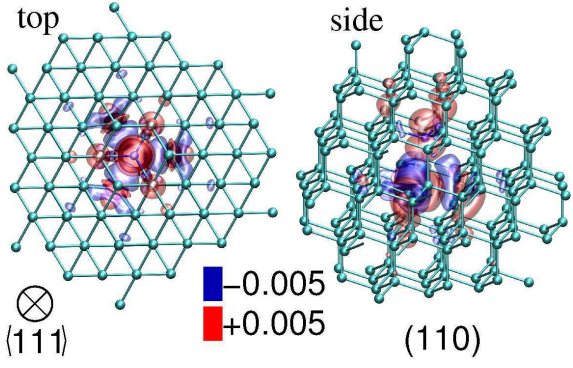}
%\end{center}
\caption{\label{fig:NVdens}(Color online) The calculated spin density
  difference between the excited $^3E$ and the ground $^3A_2$ states
  of the NV center. We choosed two representative isosurface values
  indicated in the colored rectangles. The blue(cyan) ball(s)
  represent the nitrogen(carbon) atoms. The vacant site is below the
  nitrogen atom.}
\end{figure}

In the excited state we found a significant re-arrangement of atoms
compared to the ground state. The C$_a$ atoms now farther from the
vacant site (1.70~\AA ) than the N-atom (1.63~\AA ) in contrast to the
case of the ground state. The N-atom attracts less electrons from its
neighbor C-atoms: the Bader-charge of N-atom (5.88$e$) is 0.09$e$ less
than in the ground state. Consequently, the Bader-charge of C$_b$
atoms increases by 0.03$e$. It is important to note that the
Bader-charge of C$_a$ atoms increases from 3.97$e$ to 4.04$e$. Thus,
the excitation induces a change in the dipole moment of the
defect. Next, we discuss the change in the spin density distribution
upon excitation. Fig.~\ref{fig:NVdens} shows the calculated difference
of the spin densities in the excited and ground states.
As one can see that the spin density enhanced a lot around N-atom
(indicating with red lobes) while it dropped around the C$_a$ atoms
(indicating with blue lobes). This can be explained by the hole left
on the $a_1$ defect level in the gap after excitation. The $a_1$
defect level is significantly localized on the N-atom
\cite{Gali08}. Thus, the spin polarization of the $a_1$ defect level
will spin polarize the N-atom considerably. Consequently, the spin
polarization of the C$_a$ atoms will be smaller. According to the
calculations the hyperfine constants of $^{13}$C isotopes are dropped
by around 50\% (see Table~\ref{tab:hyperf}). However, the overall
magnetization density of the N and C$_a$ atoms is about 95\% the same
\emph{both} in the ground and excited states. In other words, the spin
density mostly redistributed between the N-atom and the three
C$_a$-atoms upon excitation.   
\begin{table}
  \caption{\label{tab:hyperf} $^3A_2$ ground state (rows 2,3) versus
    $^3E$ excited state (rows 4,5). The calculated principal values of the
    hyperfine tensor (columns 2 to 4) compared to the known experimental
    data (columns 5 to 7) in MHz. The experimental data on $^{15}$N is
    taken from
    Refs.~\onlinecite{felton:075203,fuchs:117601}. 
    See text for the meaning of the question mark.}
\begin{ruledtabular}
\begin{tabular}{ccccccc}
Atom & A$_{11}$ & A$_{22}$ & A$_{33}$ & A$^\text{exp}_{11}$ &
  A$^\text{exp}_{22}$ & A$^\text{exp}_{33}$\\ \hline
 $^{15}$N          &  2.7       &     2.7      &     2.3
                   &  3.65(3)   &    3.65(3)   &  3.03(3)  \\  
      $^{13}$C(3$\times$)   &  119.7     & 120.4      &  201.1 
                            &  121.1(1)  & 121.1(1)   &  199.21(1) \\
\hline
     $^{15}$N               & -39.2   &  -39.2  & -57.8
                            & ?  &   ?  &    61$\pm$6  \\
     $^{13}$C(3$\times$)   &  56.7 &  56.7  &   126.0 &   &  &  \\
\end{tabular}
\end{ruledtabular}
\end{table}
Indeed, it has been very recently found by using excited state
spectroscopy in $^{15}$N enriched diamond samples that the $^{15}$N
hyperfine signal is $\approx$20 times larger ($\sim$60~MHz) in the
excited state than in the ground state \cite{fuchs:117601}. Our
calculations can explain this feature. However, the applied model
Hamiltonian for describing the EPR of $^{15}$N hyperfine signal
($A^\text{(N)}$) was incomplete in
Refs.~\onlinecite{fuchs:117601,jacques:057403}. Fuchs \emph{et al.}
and Jacques \emph{et al.} studied individual NV centers by confocal
photoluminescence miscroscopy where the actual defect was aligned to
the [111] axis and the applied magnetic field was parallel to this
axis, thus the angular dependence of $A^\text{(N)}$ was not
measured. They assumed isotropic $A^\text{(N)}$ for the excited state
while our study shows that it is strongly anisotropic. Fuchs \emph{et
al.} also noticed that $A^\text{(N)}$) should have the opposite sign
in the ground and excited states \cite{fuchs:117601}. Our study shows
that $A^\text{(N)}$ is positive(negative) in the ground(excited)
states in contrast to the previous assumptions
\cite{fuchs:117601,jacques:057403}.

Now, we discuss the consequence of our findings in the light of recent
experiments on the dynamic polarization of single nuclear spins of the
NV center \cite{fuchs:117601,jacques:057403}. It has been demonstrated
that the effective nuclear-spin temperature corresponds to a $\mu$K in
this process \cite{jacques:057403} \emph{decoupled} from the ambient
temperature that can be the basic physical process in the measurement
of zero-point fluctuations \cite{wrachtrup09}. In these measurements
the de-polarization of the nuclear spins of $^{15}$N
\cite{fuchs:117601,jacques:057403} and $^{13}$C$_a$
\cite{jacques:057403} have been demonstrated. This has been achieved
by the level anticrossing (LAC) of the electron spin $m_s$ sublevels
in the excited state. The LAC effect may appear if the $m_s$ sublevels
cross at a given external magnetic field (see
Fig.~\ref{fig:structlev}c,d). We show a refined and corrected model of
Ref.~\onlinecite{jacques:057403} accounted for LAC. We study the
de-polarization of $^{15}$N isotope but it can be generalized for the
$^{13}$C isotopes straightforwardly. The Hamiltonian of the system
(with neglecting the nuclear-Zeeman splitting) can be written as
\cite{jacques:057403},
\begin{equation}
 H = D_\text{es} \hat{S_z^2} + g_e \mu_B B \hat{S_z} + A_\text{es} \hat{
 \underline{S}}\, \hat{\underline{I}} \, , 
\label{eq:H}
\end{equation} 
where $\hat{\underline{S}}$ and $\hat{\underline{I}}$ are the electron
and nuclear-spin operators, $D_\text{es}$ the excited-state zero-field
splitting, $g_e$ the electron $g$ factor, $\mu_B$ the Bohr-magneton,
and $A_\text{es}$ the excited hyperfine coupling. We assume positive
$B$-field. Because $A_\text{es}$ is anisotropic, the $A_\text{es}
\hat{ \underline{S}}\, \hat{\underline{I}}$ term can be written with
the $a_\text{es}$ and $b_\text{es}$ hyperfine splittings and the
spin-shift operators as,
\begin{equation}
\frac{\hat{S_+}\,\hat{I_-} + \hat{S_-}\,\hat{I_+}}{2} \,
(a_\text{es}-b_\text{es}) + \hat{S_z}\, \hat{I_z} (a_\text{es} + 2
b_\text{es})
\label{eq:SI}
\end{equation}
The hyperfine field of $^{15}$N is parallel to the symmetry axis,
and $(a_\text{es}-b_\text{es})$=$A_{\perp}$$\approx$-39~MHz while
$(a_\text{es} + 2 b_\text{es})$=$A_{||}$$\approx$-58~MHz. According to
a recent study \cite{Batalov09} D$_\text{es}$=+1.42~GHz ($m_s = 0$
sublevel is below $m_s = \pm 1$ sublevels), so we can restrict our
study to the excited state $m_s = 0$ and $m_s = -1$ sublevels (see
Fig.~\ref{fig:structlev}d). In the basis $[
|-1,\downarrow\rangle;|-1,\uparrow\rangle;|0,\downarrow\rangle;|0,\uparrow\rangle;
]$ and by choosing the origin of energy level at level
$|0,\uparrow\rangle$, the Hamiltonian described by
Eqs.~\ref{eq:H},\ref{eq:SI} can be written as \\ $H = \left (
\begin{array}{cccc} \Et_{-1}^\downarrow - c & 0 & 0 & 0 \\ 0 &
\Et_{-1}^\uparrow - c & d & 0 \\ 0 & d & 0 & 0 \\ 0 &
0 & 0 & 0 \\
\end{array} \right )$ 
; $\begin{array}{c}\Et_{-1}^{\downarrow\uparrow}=D_\text{es}\pm
A_{||}/2 \\ c = g_e \mu_B B \\ d = A_{\perp}/\sqrt{2}\end{array}$.

%We note that $\Et_{-1}^\downarrow < \Et_{-1}^\uparrow$ because
%$A_\text{es}$ is negative for $^{15}$N. 
The eigenstates of this Hamiltonian are $|0,\uparrow\rangle$,
$|-1,\downarrow\rangle$, $|+\rangle = \alpha|0,\downarrow\rangle +
\beta |-1,\uparrow\rangle$ and $|-\rangle = \beta|0,\downarrow\rangle
- \alpha |-1,\uparrow\rangle$. By following the arguments in
Ref.~\onlinecite{jacques:057403}, the transition from the ground state
$|0,\uparrow\rangle$ to the excited state remains nuclear spin
conserving, 
%(and not the $|0,\uparrow\rangle$ with assuming
%\emph{positive} $^{15}$N hyperfine splitting in
%Ref.~\onlinecite{jacques:057403})
whereas the transition from
$|0,\downarrow\rangle$ results in $(\alpha|+\rangle + \beta|-\rangle)$
in the excited state (see Fig.~\ref{fig:structlev}e). This
superposition state then starts to precess between 
%$\alpha|+\rangle +
%\beta|-\rangle = |0,\downarrow\rangle$ and $\alpha|+\rangle -
%\beta|-\rangle = (\alpha ^2 - \beta ^2) |0,\downarrow\rangle +
%2\alpha\beta |-1,\uparrow\rangle$ 
the appropriate states at frequency $\Omega = 1/(2\hbar) \times
[(\Et_{-1}^\uparrow -c)^2 + 4d^2]^{1/2}$, where $\hbar$ is Planck's
constant. The precession frequency depends on $B$ via electron
Zeeman-effect ($c$ in our notation) that is minimal at LAC resonance,
and will be equal to $|d|/\hbar = |A_\perp|/\sqrt{2}\hbar$. Jacques
\emph{et al.} assumed isotropic hyperfine splitting for $^{15}$N,
therefore, they applied $\approx$60~MHz in this formula
\cite{jacques:057403}. Our analyzis shows that rather the
$|A_\perp|$$\approx$39~MHz should be substituted here. Nevertheless,
this precession frequency is still at the same order of magnitude as
the excited state decay rate, 12~ns \cite{fuchs:117601}. Thus, the
spin-flip process is very efficient between $|0,\downarrow\rangle$ and
$|-1,\uparrow\rangle$ states, and we can explain the de-polarization
of $^{15}$N found in the experiments
\cite{Gaebel06,fuchs:117601,jacques:057403}. It has been found that
the probability of the de-polarization effect significantly depends on
the misalignment of the magnetic field from the symmetry axis
\cite{jacques:057403}. This may be partially explained by the
anisotropy of the $^{15}$N hyperfine splitting beside the mixing of
the spin states.

We found other intriguing properties of the spin density in the
excited state. As apparent in Fig.~\ref{fig:NVdens}, the spin density
and the corresponding hyperfine tensors change considerably also for
the proximate $^{13}$C isotopes. We show the hyperfine tensors only
for the most significant change, for C$_a$ atoms in
Table~\ref{tab:hyperf}. Beside that new $^{13}$C isotope becomes
active above N-atom that has negligible spin density in the ground
state. 
%The spin-polarized N-atom can obviously polarize its neighbor
%C-atoms. 
In addition, the spin density of the sets of 6(3) C-atoms at
$R$=3.9~\AA\ decreases(increases) due to excitation where $R$ is the
distance from the vacant site. According to our previous study
\cite{Gali08}, one of these $^{13}$C isotopes was manipulated in the
qubit and quantum register applications
\cite{Childress06,Childress07}. Our study shows that during the
optical set and read-out processes the spin-density of the addressed
proximate $^{13}$C isotopes changes indicating an effective
oscillating magnetic field with the inverse lifetime of the
excited-state. This may also influence the decoherence of the
entangled electron-nuclear spin state that has not yet been considered
\cite{maze:094303}.
%, and may be taken into account
%in the analyzis of this process \cite{maze:094303}.

%\section{Summary and Conclusions}
%In summary, we showed that the spin density and the hyperfine
%splitting change significantly due to optical excitation in the
%negatively charged nitrogen-vacancy defect in diamond. Our analyzis
%showed that the previous models applied for this system should be
%refined and we could resolve some controversies regarding the
%excited-state spectroscopy of this defect. 

%\section*{Acknowledgments}
AG acknowledges support from Hungarian OTKA No.\ K-67886. The fruitful
discussion with Jeronimo Maze is appreciated.

%\bibliographystyle{prsty4}
%\bibliography{references}

\begin{thebibliography}{10}

\bibitem{Gruber97}
A. Gruber {\it et~al.}, Science {\bf 276},  2012  (1997).

\bibitem{Epstein05}
R.~J. Epstein, F. Mendoza, Y.~K. Kato, and D.~D. Awschalom, Nat. Phys. {\bf 1},
   94  (2005).

\bibitem{Childress06}
L. Childress {\it et~al.}, Science {\bf 314},  281  (2006).

\bibitem{Hanson08}
R. Hanson {\it et~al.}, Science {\bf 320},  352  (2008).

\bibitem{Maze08}
J.~R. Maze {\it et~al.}, Nature {\bf 455},  644  (2008).

\bibitem{Balasubramanian08}
G. Balasubramanian {\it et~al.}, Nature {\bf 455},  648  (2008).

\bibitem{wrachtrup09}
J. Wrachtrup, Nature Physics {\bf 5},  248  (2009).

\bibitem{Jelezko04-2}
F. Jelezko {\it et~al.}, Phys. Rev. Lett. {\bf 93},  130501  (2004).

\bibitem{Jelezko04-1}
F. Jelezko {\it et~al.}, Phys. Rev. Lett. {\bf 92},  076401  (2004).

\bibitem{Gaebel06}
T. Gaebel {\it et~al.}, Nature Physics {\bf 2},  408  (2006).

\bibitem{Childress07}
M.~V. {Gurudev Dutt} {\it et~al.}, Science {\bf 316},  312  (2007).

\bibitem{Neumann08}
P. Neumann {\it et~al.}, Science {\bf 320},  1326  (2008).

\bibitem{jacques:057403}
V. Jacques {\it et~al.}, Phys. Rev. Lett. {\bf 102},  057403  (2009).

\bibitem{maze:094303}
J.~R. Maze, J.~M. Taylor, and M.~D. Lukin, Phys. Rev. B {\bf 78},  094303
  (2008).

\bibitem{felton:075203}
S. Felton {\it et~al.}, Phys. Rev. B {\bf 79},  075203  (2009).

\bibitem{fuchs:117601}
G.~D. Fuchs {\it et~al.}, Phys. Rev. Lett. {\bf 101},  117601  (2008).

\bibitem{He93}
X.-F. He, N.~B. Manson, and P.~T.~H. Fisk, Phys. Rev. B {\bf 47},  8816
  (1993).

\bibitem{rabeau:023113}
J.~R. Rabeau {\it et~al.}, Applied Physics Letters {\bf 88},  023113  (2006).

\bibitem{Gali08}
A. Gali, M. Fyta, and E. Kaxiras, Phys. Rev. B {\bf 77},  155206  (2008).

\bibitem{Davies76}
G. Davies and M.~F. Hamer, Proc. R. Soc. London Ser. A {\bf 348},  285  (1976).

\bibitem{Goss96}
J.~P. Goss {\it et~al.}, Phys. Rev. Lett. {\bf 77},  3041  (1996).

\bibitem{Loubser77}
J.~H.~N. Loubser and J.~P. van Wyk,  in {\em Diamond Research (London)}
  (Industrial Diamond information Bureau, London, 1977), pp.\ 11--15.

\bibitem{Batalov09}
A. Batalov {\it et~al.}, arXiv:0902.2330, 2009.

\bibitem{PBE}
J.~P. Perdew, K. Burke, and M. Ernzerhof, Phys. Rev. Lett. {\bf 77},  3865
  (1996).

\bibitem{Kresse96}
G. Kresse and J. Furthm\"uller, Phys. Rev. B {\bf 54},  11169  (1996).

\bibitem{Blochl94}
P.~E. Bl\"ochl, Phys. Rev. B {\bf 50},  17953  (1994).

\bibitem{MP76}
H.~J. Monkhorst and J.~K. Pack, Phys. Rev. B {\bf 13},  5188  (1976).

\bibitem{CPPAW}
P.~E. Bl\"ochl, C.~J. F\"orst, and J. Schimpl, Bull. Mater. Sci. {\bf 26},  s33
   (2001).

\bibitem{baderanalyzis}
W. Tang, E. Sanville, and G. Henkelman, J. Phys.: Condens. Matter {\bf 21},
  084204 (7pp)  (2009).

\bibitem{Lenef96}
A. Lenef and S.~C. Rand, Phys. Rev. B {\bf 53},  13441  (1996).

\end{thebibliography}

\end{document}